\documentclass[doublecol]{epl2}

\title{Enhancing synchronization in growing networks}
\usepackage{amssymb}
\usepackage{xcolor}
\author{Y. Wang\inst{1,2}, A. Zeng\inst{3}\footnote{an.zeng@unifr.ch}, Z. Di\inst{1,2}  \and Y. Fan\inst{1,2}\footnote{yfan@bnu.edu.cn}}
\shortauthor{Y. Wang \etal}

\institute{
  \inst{1} Department of Systems Science, School of Management, Beijing Normal University, Beijing 100875, P.R. China\\
  \inst{2} Center for Complexity Research, Beijing Normal University, Beijing 100875, P.R. China\\
  \inst{3} Department of Physics, University of Fribourg, Chemin du Mus\'{e}e 3, CH-1700 Fribourg, Switzerland\\
}
\pacs{89.75.Hc}{Networks and genealogical trees}
\pacs{05.45.Xt}{Coupled oscillators}
\pacs{89.75.-k}{Complex systems}

\abstract{Most real systems are growing. In order to model the evolution of real systems, many growing network models have been proposed to reproduce some specific topology properties. As the structure strongly influences the network function, designing the function-aimed growing strategy is also a significant task with many potential applications. In this letter, we focus on synchronization in the growing networks. In order to enhance the synchronizability during the network evolution, we propose the Spectral-Based Growing (SBG) strategy. Based on the linear stability analysis of synchronization, we show that our growing mechanism yields better synchronizability than the existing topology-aimed growing strategies in both artificial and real-world networks. We also observe that there is an optimal degree of new added nodes, which means adding nodes with neither too large nor too low degree could enhance the synchronizability. Furthermore, some topology measurements are considered in the resultant networks. The results show that the degree, node betweenness centrality from SBG strategy are more homogenous than those from other growing strategies. Our work highlights the importance of the function-aimed growth of the networks and deepens our understanding of it.}

\begin{document}

\maketitle

\section{Introduction}
Networks, despite their simplicity, represent the interaction structure among components in a wide range of real complex systems, from social relationships among individuals, to interactions of proteins in biological systems, and even the interdependence of function calls in large software projects. The network concept has been developed as an important tool for analyzing the relationship of structure and function for many complex systems in the last decades~\cite{Barabasi2002,Newman2003,Barabasi1999,Watts1998,Boccaletti2006}. An interesting phenomenon observed in complex networks is synchronization~\cite{Barabasi2002}. As a kind of collective behavior, synchronization is often encountered in living systems, such as circadian rhythm, phase locking respiration with mechanical ventilator, phase locking of chicken embryo heart cells with external stimuli~\cite{Alex2008}. The early works on synchronization were concerned with only a small
number of coupled oscillators. However, as the synchronization in many real-world systems is based on a large number of dynamical units interacting with a complex coupling
structure, the research of synchronization on complex networks has attracted much attention recent years.

Previous studies have given us an overview over the synchronizability of some well-known network models. In general,
random networks have better synchronizability than regular networks, and Strogatz-Watts networks have better
synchronizability than scale-free networks~\cite{Hong2002,Wang2002,Barahona2002}. Actually, there are many factors that affect the network synchronizability. For example, average shortest distance is an important factor. Besides, the heterogeneity of the network is also
one of the most influential factors determining the synchronizability~\cite{Hong2004}. Generally speaking, the less heterogeneous the network is, the better its synchronizability will be. With these understandings, many methods have been proposed to enhance synchronizability in complex networks. There are several main directions including designing strategies for coupling strength~\cite{PRE016116,PRL034101,PRL218701,PRL138701,EPL48002,PRE056123}, modifying network structures~\cite{Donetti2005,WangB2006,Ajdari2008,Hagberg2008}, flipping the directionality~\cite{PRL228702,EPJB217,PRE045101} and so on. Each group of methods has its corresponding application field since the properties of real systems may form different operation constraints.

Most of the former works on synchronization focus on given size networks. However, most real-world networks are growing, which is
evidenced by numerous systems including man-made and
natural ones. Most of the previous growing mechanisms aim at reproducing some specific topology properties. For instance, in the pioneering work of Barab\'{a}si-Albert (BA) network~\cite{Barabasi1999}, the preferential attachment growing mechanism reproduces the power-law degree
distribution observed in many realistic complex systems. However, the growing mechanisms for network functions also need to be explored. For example, in some technological networks such as electric power grids, the neural networks and even social networks, the introduction of new nodes are supposed to enhance the network functions~\cite{Jeong2003,Vazquez2002}. In these cases, the topology properties of the network should be by-products during the improvement of the functions~\cite{PRL108701}. Therefore, it is interesting to study the strategy to enhance functions in the growing networks.

In this letter, employing synchronization as the typical function, we propose the Spectral-Based Growing (SBG) strategy. We compare the new growing mechanisms with some existing topology-aimed growing mechanisms including the Preferential Attachment (PA), Reversed Preferential Attachment (RPA), Random Attachment (RA) mechanisms. We find that the SBG strategy yields a better synchronizability in both artificial and real-world networks. Moreover, detailed study on some topology measurements of the resultant networks shows that the topology measurement such as the degree and node betweenness centrality in SBG networks are more homogeneous than those in the networks from other growing mechanisms. Besides, the SBG networks enjoy smaller clustering coefficient and average shortest path length. Interestingly, we observe that the synchronizability enhancement is strongly related to the degree of new added nodes. The synchronizability would be weaken if the degree of the new nodes are too large or too low. Actually, there is an optimal degree for each given system and the optimal degree is approximately proportional to the average degree of the original networks. Finally, in order to overcome the drawback of trapping in local optimality of SBG method when a large number of nodes are added, we tested the simulated-annealing-based SBG (SSBG) strategy. With the simulated annealing process, the SSBG strategy can jump out of the local optimality and further improve the synchronizability when a large number of nodes are added to the networks. Our work deepens our understanding of the function-aimed growth of networks and may have wide potential applications.

The rest of the paper is organized as follows. In Sec. \uppercase\expandafter{\romannumeral2}, a new growing mechanism called the SBG strategy is proposed for enhancing the synchronizability of growing networks. In Sec. \uppercase\expandafter{\romannumeral3}, we will detailedly investigate the topology of the SBG networks and the optimal degree for new nodes. The application to real-world networks is in Sec. \uppercase\expandafter{\romannumeral4}. In Sec. \uppercase\expandafter{\romannumeral5}, the SBG strategy with simulated annealing process is also tested. Finally, concluding remarks are presented in section \uppercase\expandafter{\romannumeral6}.

\section{Spectral-Based Growing (SBG) Strategy}

Let's consider a system with $N$ identical oscillators symmetrically coupled through a network. The equations of motion for the oscillator state vector $\textit{\textbf{x}}$ at each node $i$ are

\begin{equation}
\dot x_i  = F(x_i) - \sigma \sum\limits_{j = 1}^N {L_{ij} H(x_j )},
\end{equation}
where $F(x)$ determines the uncoupled oscillator dynamics of each node, $H(x)$ specifies the coupling of the vector fields, $\sigma$ is the coupling strength, and $L$ is the graph Laplacian. Specifically, $L$ is defined as $L_{ij}=-1$ if an edge exists between node $i$ and $j$, and $L_{ii}=k_i$, where $k_i$ is the degree of node $i$. Solutions of this equation are defined to be synchronized if $x_i = x_j$ for all nodes $i$ and $j$ in the network. From ref.~\cite{Barahona2002}, for many oscillatory systems the master equation is negative only in a single interval $[\alpha_1,\alpha_2]$ determined by $F$ and $H$. This implies that the network is synchronizable only when the eigenratio $\lambda_N/\lambda_2 < \alpha_2/\alpha_1$. One can conclude that the synchronizability is related to the eigenratio $r=\lambda_{N}/\lambda_{2}$, where $\lambda_N$ and $\lambda_2$ are the largest and second-smallest eigenvalues of the graph Laplacian, respectively. The smaller the eigenratio $r$ of a network is, the better its synchronizability will be. Our aim is to enhance the synchronizability when the network grows, which is equivalent to minimizing the eigenratio $r$.

For a symmetric matrix $L$, the eigenvalues of this matrix and the corresponding eigenvectors are denoted as $\lambda$ and $\textit{\textbf{v}}$ respectively. When an edge is added to or removed from the network, we denote $\Delta L$, $\Delta \lambda$ and $\Delta \textit{\textbf{v}}$ as the corresponding perturbation of the Laplacian matrix $L$, its eigenvalues and its eigenvectors. We have the following equations:
\begin{equation}
(L + \Delta L)(\textit{\textbf{v}} + \Delta \textit{\textbf{v}}) = (\lambda  + \Delta \lambda )(\textit{\textbf{v}} + \Delta \textit{\textbf{v}}).
\end{equation}

Left multiplying (2) by $\textit{\textbf{v}}^T$ and neglecting second order terms $\textit{\textbf{v}}^T \Delta L\delta \textit{\textbf{v}}$ and $\textit{\textbf{v}}^T \Delta \lambda \delta \textit{\textbf{v}}$, we obtain
$
\Delta \lambda _i  \approx  \frac{{v_i^T \Delta Lv_i }}{{v_i^T v_i }}.
$
If an edge between node $m$ and $n$ is added to the network, $\Delta L_{mm} = \Delta L_{nn} = 1$ and $\Delta L_{mn} = \Delta L_{nm}= -1$. Then $\lambda_i$ increases by $(v_{i,m}  - v_{i,n} )^2$ if the edge is added to the network, where $\textit{\textbf{v}}_i$ is the eigenvector with unit norm corresponding to $\lambda_i$. This gives the first-order approximation of the increase in $\lambda_i$.

When an edge is going to be removed from the network, if $
\frac{{\lambda _N  - \Delta \lambda _{N,a} }}{{\lambda _2  - \Delta \lambda _{2,a} }} < \frac{{\lambda _N  - \Delta \lambda _{N,b} }}{{\lambda _2  - \Delta \lambda _{2,b} }}$ is satisfied, we choose edge $a$ instead of edge $b$ since the eigenratio after removing edge $a$ is smaller. Neglecting the second-order terms, we obtain that an existing edge $a$ which minimizes
$\lambda _N \Delta \lambda _{2,a}  - \lambda _2 \Delta \lambda _{N,a}$ should be removed and a nonexisting edge $a$ which maximizes $\lambda _N \Delta \lambda _{2,a}  - \lambda _2 \Delta \lambda _{N,a}$ should be added. Accordingly, we propose the Spectral-Based Growing Strategy (SBG) as follows:

1. A new node $i$ with $k$ links is added, and these $k$ links are connected randomly to the original network.

2. For each existing edge connecting node $i$ and node $j$ of the network, the quantity $C =
\lambda _N (v_{2,i} - v_{2,j})^2  - \lambda _2 (v_{N,i} - v_{N,j})^2$ is calculated. The edge with minimum $C$ is cut.

3. For each pair of unconnected nodes $i$ and $l$ of the network, the quantity $E =
\lambda _N (v_{2,i} - v_{2,l})^2  - \lambda _2 (v_{N,i} - v_{N,l})^2$ is calculated. Then an edge between a pair of nodes with maximum $E$ is created.

4. If the eigenratio $\lambda_N/\lambda_2$ of the new network becomes smaller after steps 2 and 3, the rewiring is accepted, otherwise the process is rejected.

5. Once step 4 is rejected, all the links for this new node $i$ are fixed.

We compare the SBG strategy with some topology-aimed growing strategies including the preferential attachment (PA), reversed preferential attachment (RPA) and Random attachment (RA) strategies. In PA, the probability of a node $i$ in the network connected by the new node is proportional to $k_i$, where $k_i$ is the degree of node $i$. This growing mechanism yields the power-law degree distribution observed in many realistic systems. In RPA, the probability of a node $i$ in the network connected by the new node is proportional to $k_i^{-1}$. Consequently, the node degree will be homogenous. In RA, the new nodes randomly connect to existing nodes in the original network.

\begin{figure}
\center
\includegraphics[height=7cm,width=8.8cm]{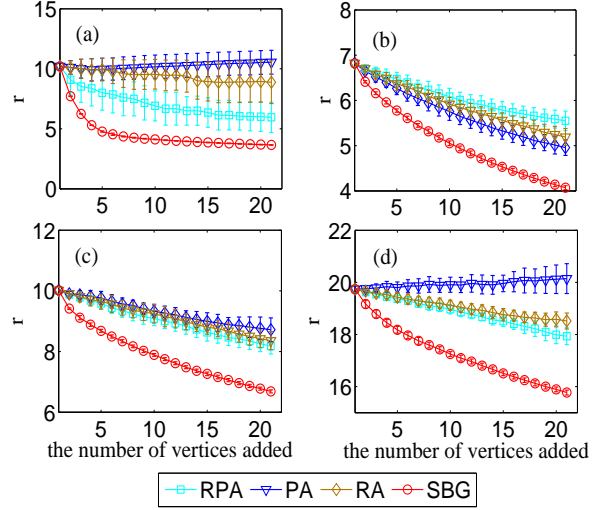}
\caption{(color on-line) Four growing mechanisms in different network models including (a) ER random graph with $\overline{k}=15$, (b) GN benchmark with $k_{in} = 12$, (c) WS small-world network with $ p = 0.01$ and $\overline{k}=20$, (d)BA network with $\overline{k}=10$. The initial network size in these four models is 300. The results are obtained by averaging over $20$ independent realizations.}\label{Fig1}
\end{figure}

In the following simulation, we mainly consider four different kinds of network models: Barab\'{a}si-Albert (BA) network with $\overline{k} = 10$~\cite{Barabasi1999}; Watts-Strogatz (WS) small-world network with $ p = 0.01$ and $ \overline{k} = 20$~\cite{Watts1998}; Erd\"{o}s-R\'{e}nyi (ER) random graph with $\overline{k}=15$~\cite{PM290}; and Girvan-Newman(GN) benchmark with $k_{in} = 12$ and $\overline{k}=16$~\cite{PRE026113}. The initial number of nodes in BA, SW and ER modeled networks is $300$, and the size of GN benchmark is 128. When a new node is introduced, the degree of it is equal to the average degree of the original network. Fig.~\ref{Fig1} shows the eigenratio $r$ as a function of the number of nodes added. Obviously, the eigenratio in SBG strategy decreases fastest in all these four network models. The eigenratio in RPA strategy performs second best because it yields a homogeneous degree distribution which is favorable for synchronization. As a widely used growing strategy, the PA strategy cannot lead to a good synchronizability since heterogenous degree distribution is formed by the PA strategy. Our result indicates that the strategy aiming at reproducing topology properties may not guarantee a high-performance of some typical function.

\section{Topology of the SBG networks and the optimal degree for new nodes}

In this section, we will detailedly discuss topology properties of the SBG networks and the effect of new nodes' degree on the resultant synchronizability.

First, we investigate the topology properties of the networks after the growth including degree, clustering coefficient, average shortest path length, node betweenness centrality. Starting from the initial networks with $300$ nodes, $20$ new nodes are added and the degree of them is equal to the average degree of the original network. We compare SBG networks with the initial, PA, RPA, and RA networks and the results are shown in table~\ref{tab1}. From the results, we observe that the standard deviation of node degree($VAR_k$), the average clustering coefficient($<C>$), the average shortest path length($<d>$), the standard deviation of node betweenness centrality($VAR_b$) and the maximum node betweenness centrality($MAX_b$) from the SBG strategy are generally smaller than those from the topology-aimed growing strategies. It has been pointed out that networks with optimal synchronizability should be a class of entangled networks with homogenous topology properties, such as degree, node betweenness centrality~\cite{Donetti2005}. From the results in table ~\ref{tab1}, we can see that the SBG networks are closest to the optimal state among all these growing strategies.

\begin{table*}
\caption{The topology properties in the resultant networks from different growing mechanisms.}
\label{tab1}
\begin{center}
\begin{tabular}{p{2cm} p{1.5cm} p{1.5cm} p{1.5cm} p{1.5cm} p{1.5cm} p{1.5cm}}
\hline
\hline
Growing Mechanism & $VAR_k$ & $<C>$ & $<d>$ & $MAX_b$ & $VAR_b$& Original Network\\
\hline
 SBG & \textbf{3.161}&\textbf{0.051} & \textbf{2.401} & \textbf{1112.7}& \textbf{184.04} &  \\
RPA  & 3.568  &0.055 & 2.403 &1266.3 & 207.23 & \\
 PA  & 3.910  &0.055 & 2.402 &1315.6 & 222.13& ER \\
 RA  & 3.736 &0.055&2.403&1301.7&210.45&\\
~\\
SBG & 7.710 & 0.076& 2.611 & \textbf{5668}& 887.17 &  \\
RPA & \textbf{7.600} & 0.075& 2.621 &5912 & \textbf{882.30} &  \\
PA & 7.950&\textbf{0.072} & \textbf{2.601} &6434 & 944.26 & BA        \\
RA & 7.870 & 0.078& 2.612 &6579 & 921.53 &          \\
~\\
SBG &\textbf{2.441}& \textbf{0.180}&\textbf{2.010}&\textbf{274.8} &\textbf{42.51}&\\
RPA&3.060&0.190 &2.021&321.9 &61.48&\\
PA&3.860&0.200 &2.033&403.7 &75.55&GN\\
RA&3.310&0.182 &2.014&405.8 &67.10&\\
~\\
SBG &\textbf{1.550}&\textbf{0.350}&\textbf{2.680}&2091.1 &\textbf{273.3}&\\
RPA&1.780&0.358 &2.690&\textbf{2000.0} &298.1&\\
PA&2.017&0.364 &2.700&2392.8 &322.1&SW\\
RA&1.847&0.361 &2.700&2338.5 &310.4&\\
\hline
\hline
\end{tabular}
\vspace*{0.0cm}
\end{center}
\end{table*}

\begin{figure}
\center
\includegraphics[height=7cm,width=8.8cm]{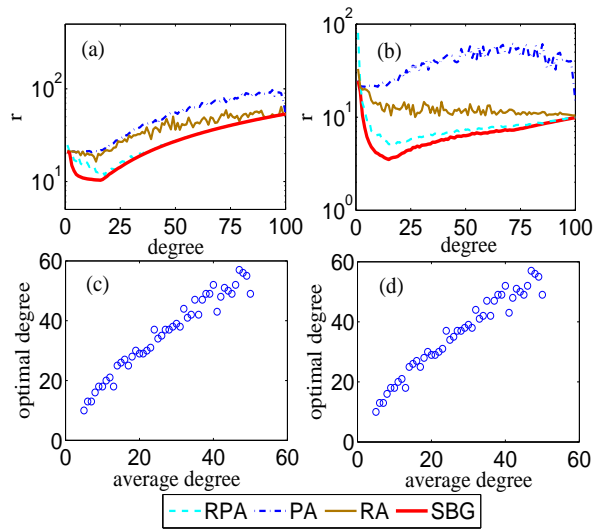}
\caption{(color on-line) The eigenratio as a function of the degree of new added nodes. (a) One node is added to the network. (b) Ten nodes are added to the network. (c) and (d) show the optimal degree in the SBG method as a function of the average degree of the original networks. The initial networks are ER models with $N=100$. In (a) and (b), the degree of initial networks is $\overline{k}=10$. The results are obtained by averaging over $20$ independent realizations.}\label{Fig2}
\end{figure}

Second, we find that the synchronizability enhancement is strongly related to the degree of new added nodes. If the degree of new added node is too small or too large, the synchronizability cannot be enhanced. Consider a ER random graph with the size of the network $N = 100$ and average degree $\overline{k}=5$, the eigenratio of it is $r=17.96$. If a new node is introduced with node degree $k = 1$, the exhaustive search method finds that the smallest eigenratio is $r=18.01$ which is even larger than that in the original network. If a new node is added with degree $k = 100$, the eigenratio is $r=55.365$ which is also larger than that in the original network. That is to say, if the degree of new added nodes is too small (e.g. $k=1$) or too large (e.g. $k=N$), the synchronizability would be weaken. Therefore, there should be an optimal degree for the new added nodes. We consider two situations in which one node and ten nodes are added to ER random networks, respectively. Fig.~\ref{Fig2}(a) and (b) shows the eigenratio as the function of the node degree of new added nodes. The results indicate that there is an optimal degree which minimizes the synchronizability of the resultant networks. In Fig.~\ref{Fig2}(c) and (d), we can see that the optimal degree of new added nodes is approximately proportional to the average degree of the original ER networks when using SBG method. Moreover, we remark that the similar relation is observed from the simulations on SW and BA models.

\section{The application to real-world networks}

In this section, we will apply the SBG strategy to some real-world networks. We mainly consider four different kinds of networks including neural, metabolic, power grid and social networks. Actually, synchronization has been shown to be of special relevance in these real systems~\cite{Alex2008}. Specifically, large-scale synchronization of oscillatory neural activity is believed to play a crucial role in the information and cognitive processing. In the metabolic networks, synchronization is important for chemical reactions of metabolism as well as the regulatory interactions that guide these reactions. Synchronization of the power grids is understood as every station and piece of equipment running on the same clock, which is crucial for its proper operation. Cascading failures related to de-synchronization can lead to massive power blackouts. In social systems, synchronization is of widely interested and related to many collective processes, such as opinion formation.

Here, we employed C.elegans neural and metabolic networks, power grid of western US and the collaboration network among scientists working at the Santa Fe Institute (SFI network)~\cite{networkdata}. We adopted the largest connected component of these networks and consider all of them as undirected ones. The neural network of C. elegans contains $302$ neurons and $2,359$ links with nodes representing neurons and links representing synaptic connections. In the metabolic network, nodes represent the chemicals produced and consumed by the reactions. If a pair of chemicals are in the same reaction, then there is a link between them. C.elegans metabolic network contains $453$ vertices and $4,596$ edges. The power grid is the network of high-voltage transmission lines that provide long-distance transport of electric power in western US. The power grid network here is with $4,994$ nodes and $6,594
$ links. The scientist collaboration network represents the collaboration relations among scientists working at the Santa Fe Institute. Edges are placed between scientists who have published at least one paper together. In SFI network, there are $118$ nodes and $200$ links.

\begin{figure}
\center
\includegraphics[height=7cm,width=8.8cm]{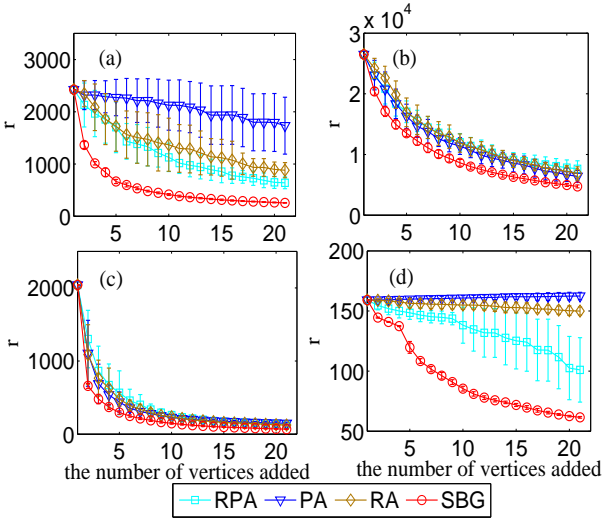}
\caption{(color on-line) The eigenratio as a function of the degree of new added nodes when SBG strategy is applied to real-world networks. The real-world networks includes (a) C.elegans metabolic network; (b) power grid; (c) SFI network; (d) C.elegans neural network. New nodes are introduced with degree equal to the average degree of the original networks. The results are obtained by averaging over $20$ independent realizations.}\label{Fig4}
\end{figure}

When new chemicals in metabolic networks, new neurons in neuron networks, new power plants in power grid or new comers in social systems are introduced, how to enhance the network synchronizability is an important problem which is meaningful in both theoretical and practical senses. Based on these real-world networks, twenty
new nodes are introduced with degree equal to the average degree of the original networks. Fig.~\ref{Fig4} shows the eigenratio $r$ as a function of the number of new nodes. As expected, we find that the SBG method has advantage in enhancing synchronizability compared to these topology-aimed methods in all the real-world networks. Moreover, the errorbar of the synchronizability from the SBG strategy is smaller than that from other methods. That is to say, our method is more effective and reliable. Just like our simulation on artificial networks, we also observe that RPA strategy leads to a good synchronizability among all the topology-aimed strategy. This result proves again the importance of homogeneous topology for synchronization when the network is growing.

\section{The SBG strategy with simulated annealing process}

When just a few nodes are added to a given network, SBG is an effective growing mechanism to enhance the network synchronizability. However, the SBG strategy can be sometimes trapped in the local maximum, especially when many nodes are added. For example, when the initial networks is with only $20$ nodes, after adding $500$ nodes the synchronizability from SBG strategy is similar to that from RPA strategy. In this section, we tested an extended SBG strategy which is called simulated-annealing-based SBG (SSBG) strategy. With the simulated annealing process, the SSBG strategy can jump out of the local maximum and further improve the SBG method when a large number of nodes are added to the networks. The algorithm consists of the following steps:

1. A new node $i$ with $k$ links is added, and these $k$ links are connected randomly to the original network.

2. For each existing edge connecting node $i$ and node $j$ of the network, the quantity $C =
\lambda _N (v_{2,i} - v_{2,j})^2  - \lambda _2 (v_{N,i} - v_{N,j})^2$ is calculated. The probability for removing an edge is proportional to $e^{\frac{1}{C}}$.

3. For each pair of unconnected nodes $i$ and $l$ of the network, the quantity $E =
\lambda _N (v_{2,i} - v_{2,l})^2  - \lambda _2 (v_{N,i} - v_{N,l})^2$ is calculated. The probability for creating an edge between the unconnected nodes is proportional to $e^{E}$.

4. If the eigenratio $\lambda_N/\lambda_2$ of the new network becomes smaller after steps 2 and 3, the rewiring is accepted, otherwise the rewiring is accepted with the probability of $
 e^{( - {{({{(\lambda _N } \mathord{\left/
 {\vphantom {{(\lambda _N } {\lambda _2 }}} \right.
 \kern-\nulldelimiterspace} {\lambda _2 }})_{New}  - {{(\lambda _N } \mathord{\left/
 {\vphantom {{(\lambda _N } {\lambda _2 }}} \right.
 \kern-\nulldelimiterspace} {\lambda _2 }})_{Old} )} \mathord{\left/
 {\vphantom {{({{(\lambda _N } \mathord{\left/
 {\vphantom {{(\lambda _N } {\lambda _2 }}} \right.
 \kern-\nulldelimiterspace} {\lambda _2 }})_{New}  - {{(\lambda _N } \mathord{\left/
 {\vphantom {{(\lambda _N } {\lambda _2 }}} \right.
 \kern-\nulldelimiterspace} {\lambda _2 }})_{Old} )} {THR}}} \right.
 \kern-\nulldelimiterspace} {THR}})}
$. Here, we fixed $THR = 3$.

5. The process is stopped after $150$ times of iterations.

\begin{figure}
\center
\includegraphics[height=5cm,width=6cm]{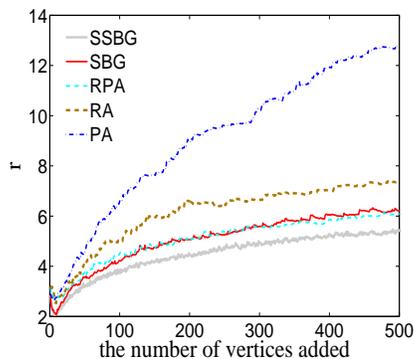}
\caption{(color on-line) The eigenratio $\lambda_N/\lambda_2$ as a function of the number of new nodes added. The initial network is an ER random graph with $N=20$, $\overline{k}=10$. New nodes are introduced with degree equal to the average degree of the original networks.}\label{Fig5}
\end{figure}

In our simulation, we start from a small ER random graph with $N = 20$, $\overline{k}=10$. Five hundred new nodes are introduced one by one. The degree of new added nodes is equal to the average degree of the original networks. Consequently, the resulting network is independent of the initial network since the size of the initial network is sufficiently small. Fig.~\ref{Fig5} shows eigenratio $r$ as a function of the number of new nodes added. The results indicates that SSBG method indeed yields a network with better synchronizability than the topology-aimed methods.

Finally, from the computational point of view, the SSBG method is with higher computational complexity than the original SBG method. However, compared with the ordinary simulated-annealing method which directly enhances synchronizability by randomly moving one link in each iteration, SSBG method can significantly reduce the computation cost since steps 2 and 3 can speed up the convergence of the whole process.

\section{Conclusion}

In this letter, aiming at improving the synchronizability in the growing networks, we propose the Spectral-Based Growing (SBG) strategy. We compare the SBG strategy with some existing topology-aimed growing ones including the Preferential Attachment (PA), Reversed Preferential Attachment (RPA), Random Attachment (RA) mechanisms. The results show that the SBG strategy yields a better synchronizability in both artificial and real-world networks.

Moreover, some topology measurements are considered in the resultant networks. It shows that the topology measurement such as degree and node betweenness centrality in SBG networks are more homogeneous than those in the networks from the other growing mechanisms. Interestingly, we find that the synchronizability enhancement is strongly related to the degree of new added nodes. The synchronizability would be weaken if the degree of the new nodes are too large or too low. Actually, there is an optimal degree for each given system and the optimal degree is approximately proportional to the average degree of the original networks.

Finally, in order to overcome the drawback of trapping in local maximum of SBG method when a large number of nodes are added, we tested the simulated-annealing-based SBG (SSBG) strategy. With the simulated annealing process, the SSBG strategy can jump out of the local maximum and further improve the SBG method when a large number of nodes are added to the networks. All these findings highlight the importance of the functional growth of networks and deepen our understanding of it.

\acknowledgments
This work is supported by the NSFC under grants No. 60974084, NCET-09-0228, and fundamental research funds for the Central Universities of Beijing Normal University.

\end{document}